\documentclass{epl}

\def\maj#1{\ifmmode\mbox{\usefont{U}{msb}{m}{n}#1}\else{\usefont{U}{msb}{m}{n}#1}\fi}
\def\v#1{\mathbf{#1}}

\usepackage{graphics,epsfig,graphicx}
\usepackage{color}

\title{Electron teleportation between quantum dots using virtual dark exciton}
\author{M. Combescot, O. Betbeder-Matibet, V. Voliotis}
\institute{INSP, Universit\'e Pierre et Marie Curie-Paris 6,
Universit\'{e} Denis Diderot-Paris 7, CNRS, UMR 7588, Campus Boucicaut, 140 rue
de Lourmel, 75015 Paris, France}
\begin{document}

\maketitle

\begin{abstract}
We here propose a mechanism to teleport electrons between quantum dots through
the transformation of a virtual bright exciton into a dark exciton. This
mechanism relies on the interactions of two composite bosons: a pair of
electrons with opposite spins, trapped in two dots and an electron-hole pair in
a free exciton coupled to an unabsorbed pump pulse, which makes it ``bright''
but virtual. This bright exciton first turns ``dark'' by dropping its electron
and stealing the trapped electron with opposite spin through an
exchange Coulomb
process with the trapped pair. In a second step, the dark exciton
``flies'' with
its electron to the other dot where it turns bright again, by the
inverse process.
The ``Shiva diagrams'' for composite boson many-body effects that we
have recently introduced, enlighten this understanding.
\end{abstract}

Spin degrees of freedom as qubits are under intensive studies due to
their potential applications in spintronics and quantum computing.
Ultrafast optical control of exciton spin in single quantum dots [1] and
its coherent control in two dots [2] through electron-hole exchange, \emph{i.\
e.}, valence-conduction transition [3], have been recently performed.
However, to
transfer excitons between dots just amounts to transfer an excitation, each
electron staying in its dot. To transfer single electrons, not electron-hole
pair, is far more a challenge.

We here propose a physical mechanism to
teleport a single electron between two semiconductor quantum dots, through the
transformation of a virtual bright exciton coupled to an unabsorbed pump pulse,
into a dark exciton.

In a 1986 pioneer experiment, Dani\`{e}le Hulin and coworkers [4] have shown
that all photons, including the unabsorbed ones, act on semiconductors
--- as proven by a blue shift of the exciton line. This shift, which
disappears when the unabsorbed pump is off, provides a physical mechanism
for ultrafast optical gates. We explained this exciton optical Stark
effect [5,6]
by the interactions of two composite bosons: the real exciton produced by the
absorbed probe photon and a virtual exciton coupled to the unabsorbed
pump pulse.

We here show that similar interactions allow to transfer electrons
between quantum dots, the probe exciton being here replaced by a pair of
opposite spin electrons trapped in two dots. The transfer is insured by
the transformation of a virtual bright exciton, coupled to the unabsorbed
pump, into a virtual dark exciton, through exchange interactions with the
trapped  electrons (see Fig.1). These interactions give rise to
a singlet-triplet splitting of the electron pair. The resulting entanglement of
the  trapped electron states $|+1/2,-1/2\rangle$ and $|-1/2,+1/2\rangle$
allows to teleport a single electron from one dot to the
other when the virtual excitons are present,
\emph{i.\ e.}, when the pump is on. So that this transfer can be optically
monitored by ultrafast pulses.

For simplicity, we here consider semiconductor with heavy holes only
$(\pm3/2$ spin$)$ and unabsorbed photons with a $\sigma_+$ polarization
$(S_z=+1)$. The virtual free excitons to which these photons are predominantly
coupled, thus have a
$(+3/2)$ hole and a $(-1/2)$ electron, their center-of-mass momentum being the
photon one, $\v Q_p\simeq\v 0$. After interactions with the trapped
electron pair having a $(+1/2)$ electron on $\v R_1$ and a $(-1/2)$
electron on $\v R_2$, the virtual exciton recombines to restore the
unabsorbed photon. Electrons being indistinguishable, the $(-1/2)$
electron which disappears can be the one of the virtual ``in'' exciton or
the one of the dot $\v R_2$. Let us recall that interesting nonlinearities
usually come from processes in which the virtual electron which is created
differs from the one which recombines.

\begin{figure}[t]
\centerline{\scalebox{0.3}{\includegraphics{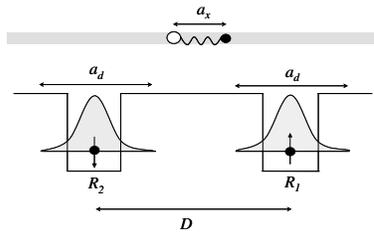}}}
\vspace{-4cm}
\caption{Free exciton coupled to unabsorbed photons,
in the presence of two electrons (full dots), with opposite spins,
trapped in two quantum dots, $R_1$ and $R_2$. Exchanges between
the electron of the delocalized exciton and the electrons trapped
in the dots make possible the electron transfer}
\end{figure}

To monitor the electron transfer optically, the dot direct coupling
through their wave function overlap, must be negligible, so that
these dots have to be reasonably far apart. The transfer we
propose is based on composite boson interactions. According to our
new theory, these interactions
generate \emph{two} scatterings [7,8,9]: the energy-like ``direct Coulomb
scatterings'' in which each boson keeps its fermions and the conceptually
novel ``Pauli scatterings'' in which the bosons exchange their fermions without
Coulomb process --- which makes them dimensionless. All the optical nonlinear
effects we have up to now studied [5,10-12] are
controlled by Pauli scatterings alone,
\emph{i.\ e.}, processes in which no Coulomb interaction takes place.
Consequently, it is not possible to describe these effects correctly through a
model Hamiltonian, whatever this model is, since Hamiltonians, by construction,
contain energy-like scatterings, while the Pauli scatterings are
dimensionless. It turns out that, in the particular case of spatially
trapped carriers, these Pauli scatterings have to be mixed with Coulomb
processes in order to produce a sizeable transfer, for rather subtle
reasons that
we now explain.

\begin{figure}[t]
\centerline{\scalebox{0.3}{\includegraphics{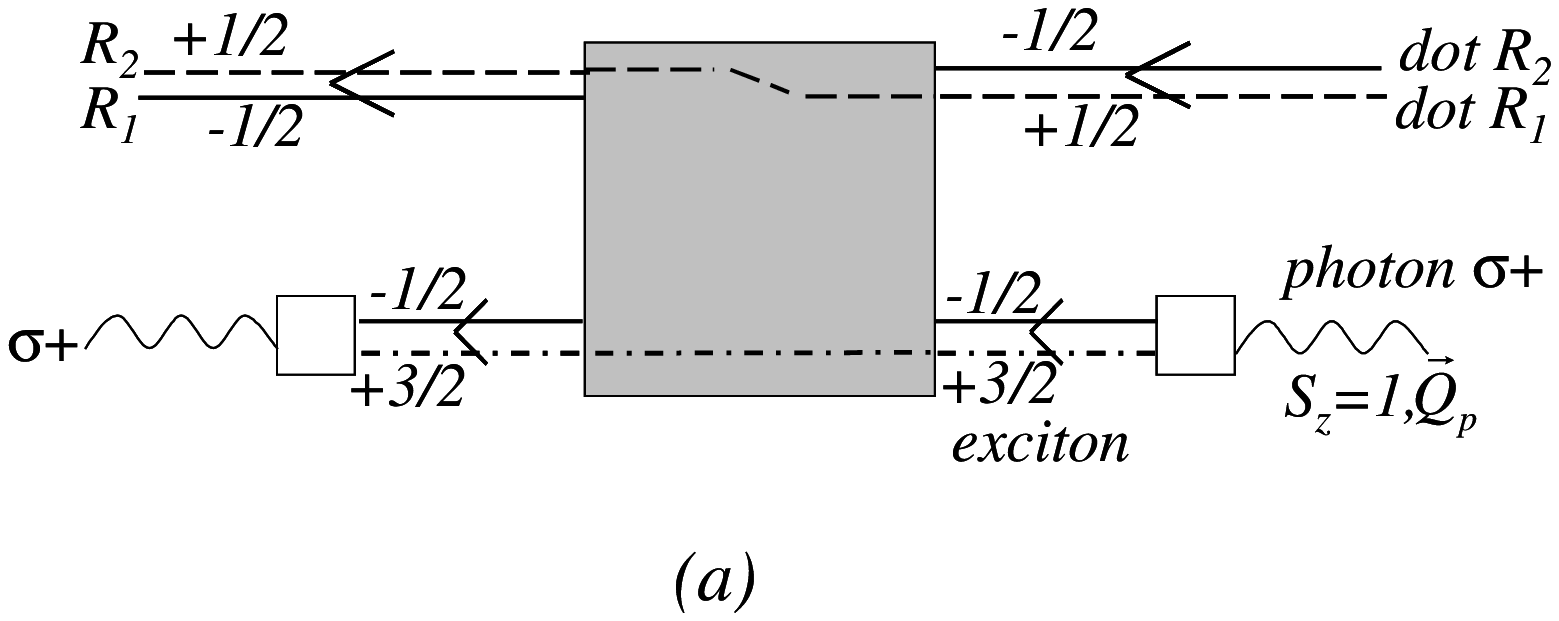}}
\scalebox{0.3}{\includegraphics{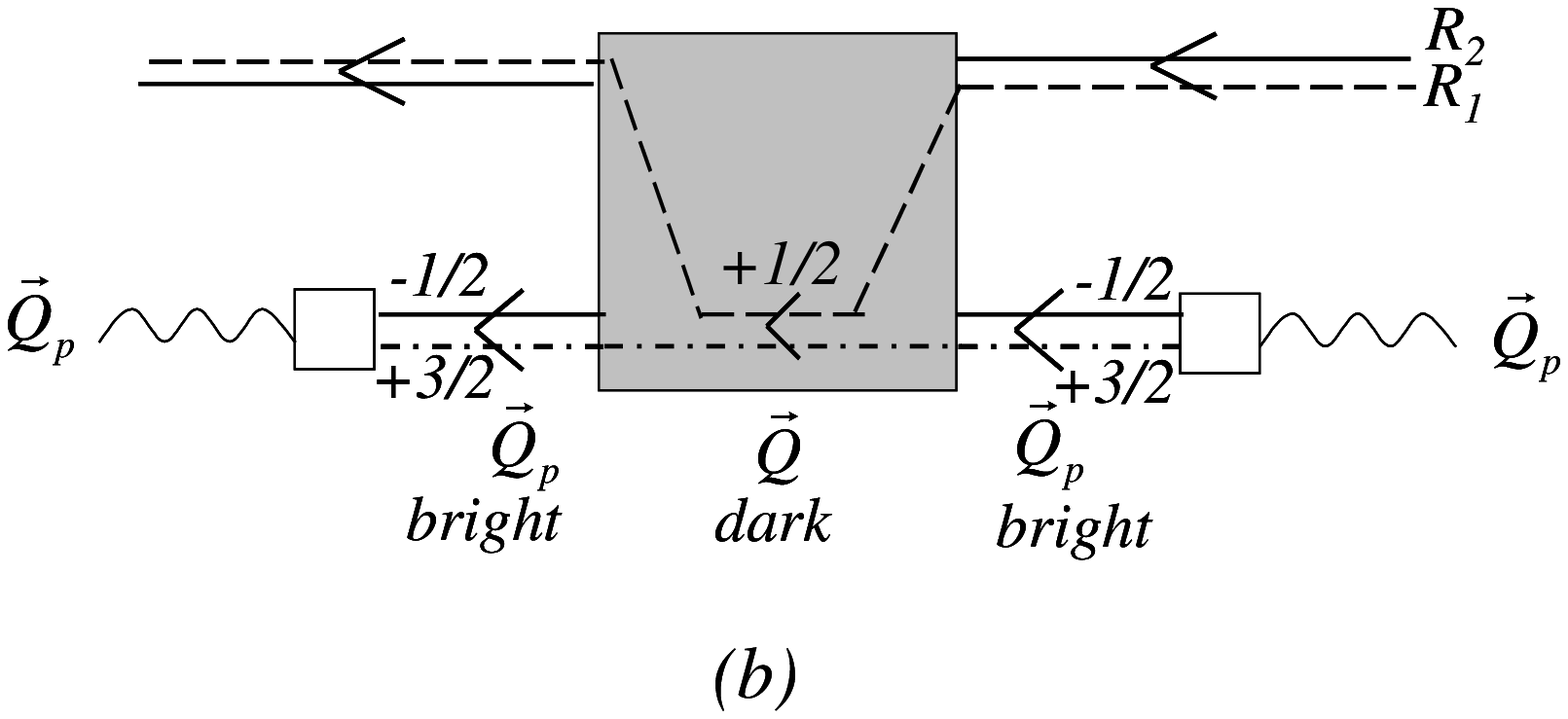}}}
 \vspace{-5cm}
\caption{In the black box, the composite boson made of the
electron-electron pair trapped in the dots, interacts with the
composite boson made of an electron-hole pair (exciton) coupled to
a $\sigma_+$ photon, in such a way that the $(+1/2)$ electron of
the dot $\v R_1$ ends on the dot $\v R_2$. This $(+1/2)$ electron
can either stay in the dots, as in (a), or join the $(+3/2)$ hole
to form a dark exciton, as in (b).}
\end{figure}

The set of interactions induced by these two scatterings can be divided
in two types:

\noindent (i) In one, the $(+1/2)$ electron stays in the dots (see
Fig.2a). As these processes contain the dot wave function overlap, they
produce a negligible transfer when the dots are far apart.

\noindent (ii) In the other, the $(+1/2)$ electron goes with the hole to form
a dark exciton (see Fig.2b). These processes contain the overlaps of the wave
function of each dot with the electron wave function in a virtual
free exciton. Since the exciton center-of-mass is delocalized over the whole
sample, these overlaps differ from zero (see Fig.1), although very small. This
is compensated by the large number of such processes, since, along with
changing its electron, the exciton also changes its momentum from $\v Q_p$ to
$\v Q$, the excess being provided by the dots.

Our new ``Shiva diagrams'' for composite boson interactions, here a trapped
electron-electron pair and a free exciton, allows to visualize the physics
of this electron transfer (see Fig.3). Dimensional arguments allow to
grasp the set of microscopic interactions
necessary to produce it:
We look for a singlet-triplet splitting, \emph{i.e.}, an energy-like
quantity. All processes linear in pump intensity
contain two energy-like semiconductor-photon couplings (see Fig.2). As the
splitting comes from composite boson interactions, it must also
contain composite boson scatterings.
Pauli scatterings being dimensionless, no energy denominators are
needed when they take place. On the opposite, each Coulomb scattering must
appear with an energy denominator, which can only be the energy difference
between the state 0, made of the trapped pair plus the photon, and one of the
various intermediate states appearing in the process at hand. Two relevant ones
are the state $\alpha$ just after the virtual exciton creation and the state
$\beta$ just before its recombination (see Fig.3). They are such that
$\mathcal{E}_0-\mathcal{E}_\alpha=\mathcal{E}_0-\mathcal{E}_\beta= -\delta$,
where
$\delta$ is the exciton-photon detuning. A third relevant denominator is
$\mathcal{E}_0-\mathcal{E}_\gamma$, where $\gamma$ is the intermediate state
having a dark exciton (see Fig.3c).

Dimensional arguments then show that the process of Fig.3a, which just contains
one Pauli scattering, can only have one energy denominator,
$\mathcal{E}_0-\mathcal{E}_\alpha$, so that it is the dominant one at large
detuning. However, as in it appears the overlap of the
$(\v R_1,\v R_2)$ dots, its contribution to the singlet-triplet
splitting  is negligible for far apart dots. Similarly, the
process of Fig.3b, which, in addition, contains one Coulomb scattering, has two
energy denominators,
$\mathcal{E}_0-\mathcal{E}_\alpha$ and $\mathcal{E}_0-\mathcal{E}_\beta$, but
the dot overlap is still present. To avoid it, the state $\gamma$ with
its dark exciton, has to appear through an energy denominator
$\mathcal{E}_0-\mathcal{E}_\gamma$, so that two Coulomb interactions are
at least needed to produce a sizeable transfer. In order to also avoid the dot
overlap for the
$(-1/2)$ electrons, we end with the virtual ``in'' exciton leaving its $(-1/2)$
electron to the dot
$\v R_1$ and the virtual ``out'' exciton having taken its $(-1/2)$
electron from
the dot $\v R_2$ (see Fig.3c).

\begin{figure}[t]
\centerline{\scalebox{0.3}{\includegraphics{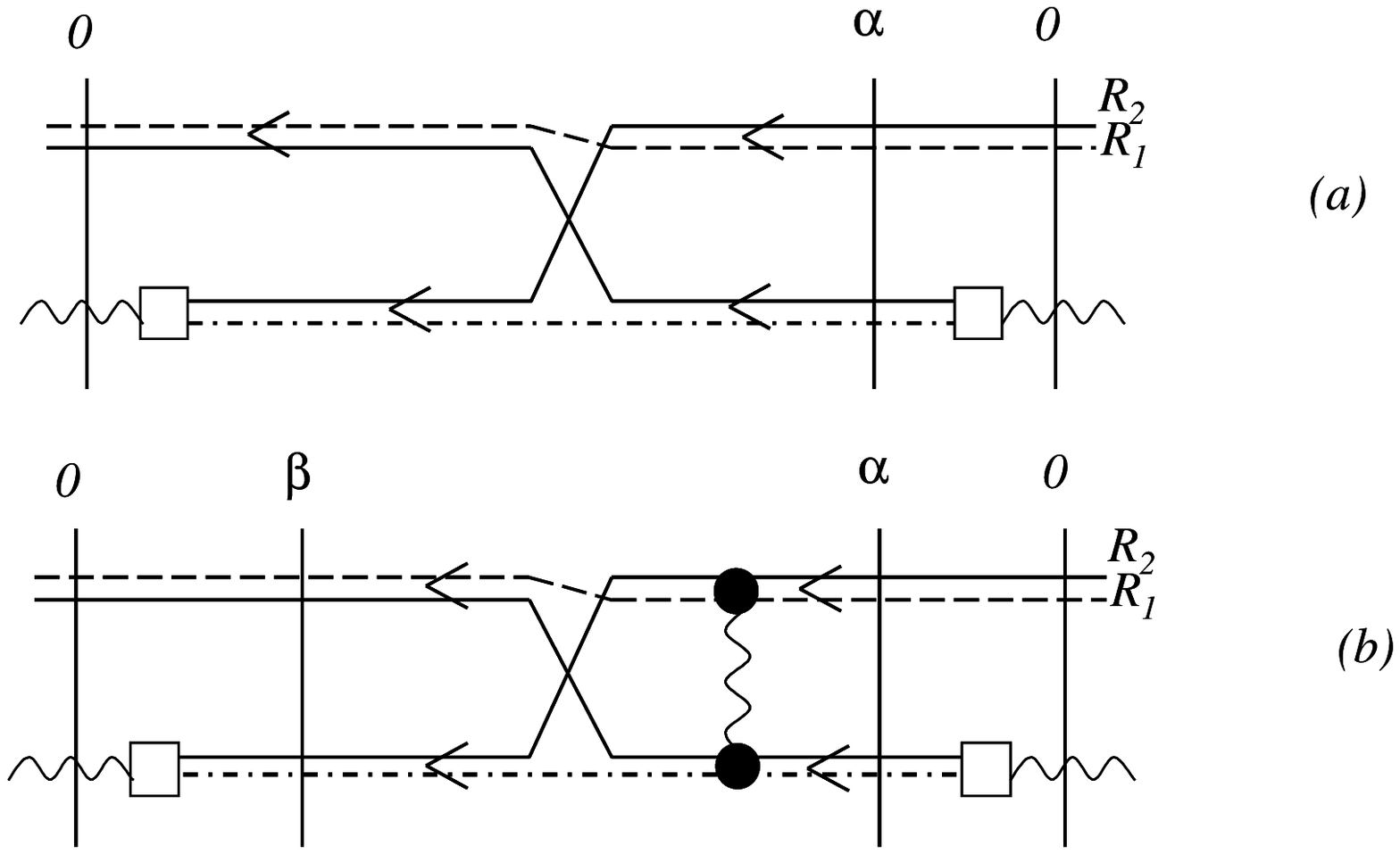}}
\scalebox{0.3}{\includegraphics{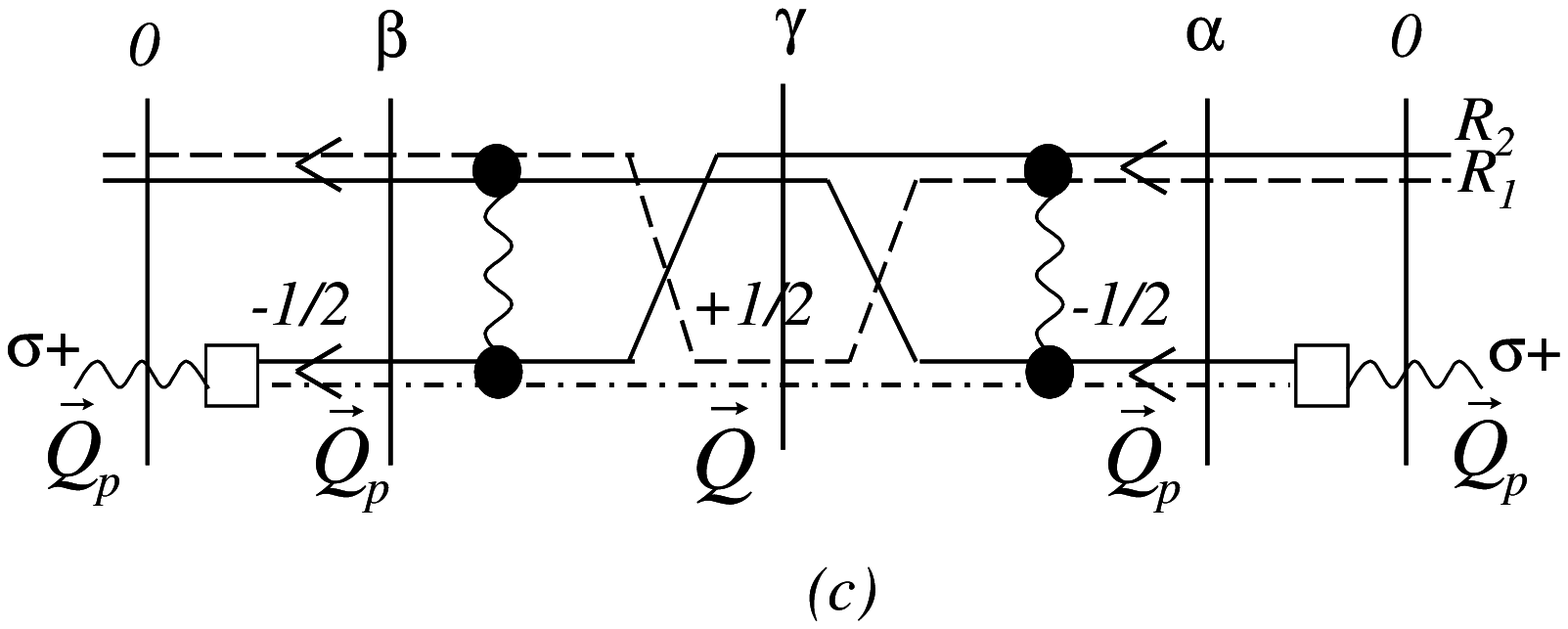}}} \vspace{-4cm}
\caption{``Shiva diagrams'' for the set of Pauli and direct
Coulomb scatterings leading to transfer the $(+1/2)$ electron from
the dot $\v R_1$ to the dot $\v R_2$. The processes (a) and (b),
having zero and one direct Coulomb scattering, make appearing the
dot wave function overlap, so that their contributions to the
singlet-triplet splitting is negligible. In (c), which contains
two Coulomb scatterings, the intermediate state $\gamma$ has a
dark exciton, while the $(-1/2)$ electron which regenerates the
unabsorbed photon $\sigma_+$, differs from the photocreated one,
as usual in optical nonlinearities.}
\end{figure}

\emph{The ``Shiva diagram'' of Fig.3c thus corresponds to the large detuning
dominant term of the singlet-triplet splitting, linear in pump
intensity}. It describes the following physics: A
$\sigma_+$ photon with momentum $\v Q_p$ creates a virtual exciton with
same momentum, made of $(+3/2)$ hole and $(-1/2)$ electron. This exciton
has a direct Coulomb scattering with the trapped pair, followed by a Pauli
scattering, \emph{i.\ e.}, a carrier exchange with the dot $\v R_1$,
in order to
become dark. In a second step, the dark exciton turns bright again by
an electron
exchange with the dot
$\v R_2$, followed by a direct Coulomb scattering. In this process, the
intermediate dark exciton thus
teleports the $(+1/2)$ electron from dot $\v R_1$ to dot $\v R_2$.

The energy denominator associated to this intermediate state $\gamma$ can
stay essentially as small as the detuning $\delta$ if the trapped
electrons and the exciton stay in their ground states. The exciton can
however change its momentum from $\v Q_p\simeq\v 0$ to $\v Q$ without too
much cost in energy if the exciton center-of-mass mass, $M_X$, is large enough.
This leads to $\mathcal{E}_0-\mathcal{E}_\gamma=-\delta-\hbar^2Q^2/2M_X$.

As shown below, the splitting of the electron
pair, which corresponds to the ``Shiva diagram'' of Fig.3c, reads as
\begin{equation}
\Delta\simeq\Omega_0\,\frac{1}{-\delta}
\sum_{\v Q}C_{\v Q}(\v
R_2;\v R_1)\,\frac{1}{-\delta-\hbar^2Q^2/2M_X}
\,C_{\v Q}^\ast(\v R_1;\v
R_2)\,\frac{1}{-\delta}\,\Omega_0^\ast\ .
\end{equation}
$\Omega_0$ is the laser coupling to the ground state exciton, $C_{\v
Q}(\v R_1;\v
R_2)$ the exchange Coulomb scattering between exciton and trapped
electron pair,
the exciton exchanging its electron with the dot $\v R_1$. Its exact value,
obtained through the microscopical procedure explained below, is given in eq.\
(11) in terms of the exciton and dot wave functions, so that it is model
dependent.

In the absence of precise experiments, let us concentrate on
qualitative behaviors with respect to the physically
relevant parameters of the problem, namely the dot distance $D$, the photon
detuning $\delta$, the free electron-hole pair Rabi energy $\Omega$,
the exciton
center-of-mass and relative motion masses, $M_X$ and $m_X$, and the dot and
exciton spatial extensions
$a_d$ and
$a_x$ (see Fig.1).
For $Q_pD\simeq
0$, it is possible to show that $C_{\v Q}(\v R_1,\v R_2)$ varies with the dot
separation as
$e^{i\v Q.D}$. This leads to a singlet-triplet splitting
which behaves as
\begin{equation}
\Delta\simeq\frac{-e^2}{a_x}\frac{\Omega^2}{\delta^2}\frac{M_X}{m_X}\,
\left(\frac{a}{a_x}\right)^d\left(\frac{a}{b_\delta}\right)^{d-2}
\left(\frac{b_\delta}{D}\right)^{\frac{d-1}{2}}e^{-D/b_\delta}\ ,
\end{equation}
where $d=(2,3)$ is the space dimension,
$b_\delta$ the ``detuning length'' defined as
$\delta=\hbar^2/2M_Xb_\delta^2$ and
$a=\mathrm{inf}(a_d,a_x)$, equation (2) being valid for rather small detunings
($b_\delta> a'=\mathrm{sup}(a_d,a_x)$). As expected for effects coming from
unabsorbed photons, a large splitting requires a small photon detuning $\delta$
and a large Rabi energy
$\Omega$, \emph{i.\ e.}, a powerful pump and a large valence-conduction
dipolar matrix element. It is also appropriate to use quantum wells
with dots as
large as the exciton (to optimize the dot-exciton overlap).

A rough estimate of the transfer time
$T\simeq \hbar/\Delta$ obtained from eq.\ (2), leads to a few hundreds of
picoseconds for reasonable values of the parameters, namely quantum wells
having two dots at
$D\simeq 100\mathrm{nm}$, and
$a_d\simeq a_x\simeq 150\AA$, so that $e^2/a_x\simeq 10\mathrm{meV}$
while $m_X/M_X\simeq 0.1$, $\Omega\simeq 0.1\mathrm{meV}$ and
$\delta\simeq 0.5\mathrm{meV}$.

Sham and coworkers [13] have suggested to use unabsorbed photons to
exchange spins between dots. However, by lack of appropriate tools to treat
interactions with composite excitons properly, their approach can only be
phenomenological: They start with a model spin-spin Hamiltonian for trapped and
delocalized electrons. This not only misses the Coulomb attraction of the
virtual hole  responsible for the exciton formation --- which essentially kills
the electron repulsion --- but also the
\emph{dimensionless} Pauli scatterings, responsible for all optical
nonlinearities. The aim of their second work [14] is to provide a
``microscopic'' derivation of the spin-spin coupling, through a two-fluid
\emph{model} for trapped and delocalized electrons. This is again highly
phenomenological, since the ``itinerant'' (or free) electron states form a
complete set for electrons, so that ``localized'' electrons, which can be
written as a sum  of ``itinerant'' electrons, do not differ from them.
Consequently, the basis they use is overcomplete. Moreover, the scatterings for
interactions between these ``two fluids'',  on which rely all their
results, are not explicitly given in ref.\ [14]: We do not see how they can
be properly derived from a really microscopic approach, \emph{i.\ e.}, an
approach just using the semiconductor Hamiltonian we use.

Let us outline our fully microscopic procedure. It relies on a
natural extension of our  many-body theory for composite excitons to composite
bosons which are not Hamiltonian eigenstates [9]. We start with the bare
semiconductor Hamiltonian,
$(H_{\mathrm{sc}}=h_e+h_h+V_{ee}+V_{hh}+V_{eh})$, and the
localization potentials
of the two dots, $(w_{\v R_1}+w_{\v R_2})$. As in ref.\ [12], we introduce the
two complete sets of creation operators for one-electron trapped states,
$\left(h_e+w_{\v R_j}-\epsilon_\mu^{(e)}\right)\,a_{\v R_j\mu s}^\dag|v
\rangle=0$ with $j=(1,2)$ and $s=(\pm 1/2)$, with $|v\rangle$ being the vacuum
state. $\mu$ characterizes the quantum level of the electron in the dot,
$\epsilon_\mu^{(e)}$ being its energy. From them, we construct the composite
boson creation operators for trapped electron pairs,
$A_n^\dag\left(^{s_2}_{s_1}
\right)=a_{\v R_1\mu_1s_1}^\dag\,a_{\v R_2\mu_2s_2}^\dag$ with
$n=(\mu_1,\mu_2)$. They are not eigenstates of the system Hamiltonian,
$H_\mathrm{sc}'=H_\mathrm{sc}+w_{\v R_1}+w_{\v R_2}$, since
\begin{equation}
\left[H_{\mathrm{sc}}'-E_n^{(ee)}\right]\,A_n^\dag\left(^{s_2}_{s_1}\right)|v
\rangle=v_n^\dag\left(^{s_2}_{s_1}\right)|v\rangle\,\neq\ 0\ ,
\end{equation}
where $E_n^{(ee)}=\epsilon_{\mu_1}^{(e)}+\epsilon_{\mu_2}^{(e)}$. The RHS of
eq.\ (3) is however small for ground state electrons, $n=0=(\mu_0,\mu_0)$, if
the dots are far apart. The other composite bosons
are the free excitons,
$\left[H_ {\mathrm{sc}}-E_i^{(X)}\right]\,B_{is_im_i}^\dag|v\rangle=0$,
the photons being predominantly coupled to them.

The scatterings between these two composite bosons appear through a set of
commutators which are a generalization [9] of those for interacting excitons
[7]. The dimensionless ``Pauli scatterings''
$\lambda_{n'i'ni}^{(ee-X)}$ are such
that
\begin{equation}
\left[D_{n'n}\left(^{s_2'\ s_2}_{s_1'\
s_1}\right)\,,B_{is_im_i}^\dag\right]=\sum\lambda_{n'i'ni}^{(ee-X)}\,
B_{i's_i'm_i'}^\dag\ ,
\end{equation}
where the ``deviation-from-boson operator'' of the electron pair $D_{n'n}$ must
be defined as
\begin{eqnarray}
\left[A_{n'}\left(^{s_2'}_{s_1'}\right),A_n^\dag\left(^{s_2}_{s_1}\right)
\right]&=&\delta_{n'n}\left(^{s_2'\ s_2}_{s_1'\ s_1}\right)
-D_{n'n}\left(^{s_2'\ s_2}_{s_1'\ s_1}\right)\ ,\nonumber\\
\delta_{n'n}\left(^{s_2'\ s_2}_{s_1'\ s_1}\right)&=&
\langle
v|A_{n'}\left(^{s_2'}_{s_1'}\right)\,A_n^\dag\left(^{s_2}_{s_1}\right)|v\rangle\
,
\end{eqnarray}
in order to have $D_{n'n}|v\rangle=0$. The ``direct Coulomb scatterings''
$\xi_{n'i'ni}^{(ee-X)}$ appear through
\begin{equation}
\left[V_n^\dag\left(^{s_2}_{s_1}\right),B_{is_im_i}^\dag\right]=
\sum\xi_{n'i'ni}^{(ee-X)}\,A_{n'}^\dag\left(^{s_2}_{s_1}\right)\,B_{i's_im_i}
^\dag\ ,
\end{equation}
where the ``Coulomb-creation potential'' of the electron pair
$V_n^\dag\left(^{s_2}_{s_1}\right)$ must be defined as
\begin{equation}
\left[H_{\mathrm{sc}}',A_n^\dag\left(^{s_2}_{s_1}\right)\right]=E_n^{(ee)}\,
A_n^\dag\left(^{s_2}_{s_1}\right)+v_n^\dag\left(^{s_2}_{s_1}\right)
+V_n^\dag\left(^{s_2}_{s_1}\right)\ ,
\end{equation}
with $v_n^\dag$ given by eq.\ (3), in order to have $V_n^\dag|v\rangle=0$:
This insures
$V_n^\dag$ to describe the interactions
of the trapped pair with the rest of the system.

As $v_0^\dag|v\rangle\simeq 0$ for far
apart dots, the four states $A_0^\dag\left(^{s_2}_{s_1}\right)|v\rangle$ with
$s=(\pm 1/2)$ are essentially degenerate in the absence of pump beam. Their
interactions with unabsorbed photons split this fourfold subspace, the changes
being obtained from the diagonalization of
$\left[W\left(\omega_p+E_0^{(ee)}
-H_{\mathrm{sc}}'\right)^{-1}W\right]$ in this
degenerate subspace, $\omega_p$ being the photon energy and $W=U+U^\dag$ the
laser-semiconductor coupling. The appropriate way to write $U$ for
$(\sigma_+,\v Q_p)$ photons, is
$U=\sum\Omega_i\,B_{i,-1/2,3/2}$, where $|\Omega_i|$ is the Rabi
energy of the exciton $i=(\nu_i,\v Q_p)$, the ground state one
$|\Omega_o|$ being the largest, by far.

The diagonalization of
$\left[W\left(\omega_p+E_0^{(ee)}-H_{\mathrm{sc}}'\right)^{-1}W\right]$ in the
twofold subspace $A_0^\dag\left(^{-s}_s\right)|v\rangle$ shows that, for
$Q_pD\simeq 0$, the $\sigma_+$ photons split these two states into a
triplet and
a singlet, according to
\begin{equation}
|\Phi_{\pm}\rangle=\frac{1}{\sqrt{2}}\left(A_0^\dag\left(^{-1/2}_{+1/2}\right)
\pm\frac{E_{-+}}{|E_{-+}|}\,A_0^\dag\left(^{+1/2}_{-1/2}\right)\right)|v\rangle\
,
\end{equation}
their energies being $\mathcal{E}_{\pm}=E_0^{(ee)}+E_{++}\pm E_{-+}$, with
\begin{equation}
E_{s's}=\langle v|U\,A_0\left(^{-s'}_{s'}\right)\,\frac{1}{\omega_p+E_0^{(ee)}
-H_{\mathrm{sc}}'}\,A_0^\dag\left(^{-s}_{s}\right)\,U^\dag|v\rangle\ .
\end{equation}
The entanglement of the two states
$A_0^\dag\left(^{-s}_{s}\right)¬v\rangle$, with $s=\pm 1/2$, resulting from the
singlet-triplet splitting $\Delta=2E_{-+}$, gives rise to a transfer time
between dots
$T=\pi\hbar/|E_{-+}|$.

$E_{s's}$ is easy to calculate within our formalism: We
first pass the $A_0^\dag$ and $A_0$ over the Hamiltonian, using eq.\ (7). As
for the exciton optical Stark effect [6], this splits $E_{s's}$ into
three terms, $\alpha_{s's}+\beta_{s's}+\gamma_{s's}$, which have zero, one and
two ``Coulomb-creation potentials''. In $\alpha_{-+}$ and
$\beta_{-+}$ which correspond to the processes of Fig.3a and Fig.3b,
appears the
overlap of the dot ground states, so that their contributions to
$E_{-+}$ is negligible.
The term with two Coulomb creation potentials $V_0^\dag$, which corresponds to
the process of Fig.3c, reads
\begin{equation}
\gamma_{-+}=\langle\psi_p|V_0\left(^{+1/2}_{-1/2}\right)\,\frac{1}{\omega_p
+E_0^{(ee)}-H_{\mathrm{sc}}'}\,V_0^\dag\left(^{-1/2}_{+1/2}\right)|\psi_p
\rangle\ ,
\end{equation}
where
$|\psi_p\rangle=\left(\omega_p-H_{\mathrm{sc}}'\right)^{-1}\,U^\dag|v\rangle$
can be approximated by $(-\delta)^{-1}\Omega_0^\ast
B_{0,-1/2,3/2}^\dag|v\rangle$, with $0=(\nu_o,\v Q_p)$, if the photon frequency
is far enough from bound exciton resonances. To calculate
$V_0^\dag\left(^{-s}_{s}\right)B_{0,-1/2,3/2}^\dag|v\rangle$, we use eq.\ (6).
This generates direct Coulomb scatterings between exciton and trapped electron
pair. However, in order to get a sizeable $\gamma_{-+}$, it is
necessary to avoid
the dot wave function overlap for both, the $(+1/2)$ and the $(-1/2)$
electrons.
So that a Pauli scattering in which the bright ``in'' exciton and the dot
$\v R_1$ exchange their electrons has to be added as well as a
Pauli scattering between the dot $\v R_2$ and the black exciton (see Fig.3c).
If we then only keep intermediate states
$\gamma$ with the smallest energy,
\emph{i.\ e.}, states in which the exciton and the trapped pair stay in
their ground states, we obtain the splitting given in eq.\ (1), where the
Coulomb-exchange scattering precisely reads
\begin{eqnarray}
C_{\v Q}^\ast(\v R_1;\v R_2)=-\int d\{\v r\}\,
\varphi_{\v R_2}^\ast(\v r_2)\,\varphi_{\v R_1}^\ast(\v
r_e)\,\phi_{\v Q}^\ast(\v
r_1,\v r_h)
\,v(\v r_1,\v r_2,\v r_e,\v r_h)\nonumber\\
\times\
\varphi_{\v R_1}(\v r_1)\,\varphi_{\v R_2}(\v r_2)\,\phi_{\v Q_p}(\v
r_e,\v r_h)\ ,
\end{eqnarray}
where
$v(\v r_1,\v r_2,\v r_e,\v r_h)=v(\v r_1-\v r_e)+v(\v r_2-\v r_e)-v(\v r_1-\v
r_h)-v(\v r_2-\v r_h)$, with $v(\v r)=e^2/r$, is the Coulomb potential
between the free exciton
and the trapped electrons in their ``in'' states. $\varphi_{\v R_j}(\v r)$ is
the trapped electron ground state wave function  and $\phi_{\v Q}(\v r_e,\v
r_h)$ the exciton ground state wave function, $\v Q$ being the
momentum gained by
the exciton in this Coulomb exchange process, the increase from $\v Q_p$ to $\v
Q$ being provided by the dots. For
$Q_pD\simeq 0$, in a size $L$ sample, a bare dimensional analysis leads to
$C_{\v Q}^\ast(\v R_1;\v R_2)\simeq (e^2/a)(a/L)^de^{i\v Q.\v D}f(Qa')$, where
$f(x)$ is a slowly decreasing function, $(a,a')$ being defined below eq.\ (2).
The splitting given in eq.\ (2) then follows from the link between exciton and
free pair Rabi energies, namely $\Omega_o=\Omega(L/a_x)^{d/2}$, and the fact
that $f(Qa')$ can be replaced by 1 for $b_\delta<a'$.

In conclusion, we propose a fully microscopic approach, free from any model
Hamiltonian, to the teleportation of electrons between quantum dots through the
transformation of a virtual bright exciton coupled to unabsorbed
photons, into a
virtual dark exciton. This approach relies on a natural extension of our
theory for composite exciton many-body effects, to composite bosons
which are not
semiconductor eigenstates. The ``Shiva diagrams'' for
$N$-body exchanges we have recently introduced, make transparent the physics
involved.


\begin{thebibliography}{99}

\bibitem{1}
T. Unold, K. Mueller, C. Lienau, T. Elsaesser, A.D. Wieck, Phys.\ Rev.\
Lett.\ \textbf{92}, 157401 (2004)

\bibitem{2}
T. Unold, K. Mueller, C. Lienau, T. Elsaesser, A.D. Wieck, Phys.\ Rev.\
Lett.\ \textbf{94}, 137404 (2005)

\bibitem{3}
A. Nazir, B. Lovett, S.D. Barrett, T.P. Spiller, G.A.D. Briggs, Phys.\ Rev.\
Lett.\ \textbf{93}, 150502 (2004)

\bibitem{4}
A. Mysyrowicz, D. Hulin, A. Antonetti, A. Migus, H. Morko\c{c}, Phys.\ Rev.\
Lett.\ \textbf{56}, 2748 (1986)

\bibitem{5}
M. Combescot, R. Combescot, Phys.\ Rev.\ Lett.\ \textbf{61}, 117 (1988)

\bibitem{6}
For a review, see M. Combescot, Phys.\ Rep.\ \textbf{221}, 167 (1992)

\bibitem{7}
M. Combescot, O. Betbeder-Matibet, Europhys.\ Lett.\ \textbf{58}, 87 (2002)

\bibitem{8}
For a short review, see M. Combescot, O. Betbeder-Matibet, Solid State Comm.\
\textbf{134}, 11 (2005), and references therein.

\bibitem{9}
M. Combescot, O. Betbeder-Matibet, Eur.\ Phys.\ J.\ B \textbf{48}, 469 (2005)

\bibitem{10}
M. Combescot, O. Betbeder-Matibet, Solid State Comm.\ \textbf{132}, 129 (2004)

\bibitem{11}
M. Combescot, O. Betbeder-Matibet, K. Cho, H. Ajiki, Europhys.\ Lett.\
\textbf{72}, 618 (2005)

\bibitem{12}
M. Combescot, O. Betbeder-Matibet, Cond-mat/0505746, submitted to
Phys.\ Rev.\ B

\bibitem{13}
C. Piermarocchi, C. Chen, L.J. Sham, D.J. Steel, Phys.\ Rev.\ Lett.\
\textbf{89}, 167402 (2002)

\bibitem{14}
G. Ramon, Y. Lyanda-Geller, T.L. Reinecke, L.J. Sham, Phys.\ Rev.\ B
\textbf{71}, 121305(R) (2005)



\end{thebibliography}
\end{document}